\documentclass[twocolumn,aps,pre,showpacs]{revtex4}

\begin{document}


\title{Charge and Salt Driven Reentrant  Order-Disorder  and
Gas-Solid Transitions in  Charged  Colloids}

\author{P.S. Mohanty  and  B.V.R. Tata$^\ast$}
\affiliation{Materials Science Division, Indira Gandhi Centre for
Atomic Research, Kalpakkam - 603 102,Tamil Nadu India}
\address{%
\begin{minipage}[t]{6.0in}
\begin{abstract}
{Monte Carlo simulations have been performed for aqueous charged
colloidal suspensions as a function of charge density($\sigma$)
on the particles  and salt concentration $C_s$. We vary the
charge density in our simulations over a range where a reentrant
solid-liquid transition in suspensions of silica and polymer
latex particles has been reported by Yamanaka {\em et al.}
[Phys.. Rev.  Lett., {\bf 80} 5806 (1998)].  We show that at low
ionic strengths a homogeneous liquid-like ordered suspension
undergoes crystallization upon increasing $\sigma$. Further
increase in $\sigma$ resulted once again a disordered state
which is in agreement with experimental observations. In
addition to this reentrant order-disorder transition, we observe
an inhomogeneous to homogeneous transition in  our simulations
when  salt is added to the disordered inhomogeneous state.  This
inhomogeneous to homogeneous disordered transition is analogous
to the solid-gas transition of  atomic systems and has not yet
been observed  in charged colloids.  The reported experimental
observations on charged colloidal suspensions are discussed in
the light of present simulation results.}
\end{abstract}
\end{minipage}}
\maketitle

\section{INTRODUCTION}

Monodisperse charge  stabilized colloidal suspensions are
studied with great interest because they exhibit a variety of
structural orderings  analogous to that occur in atomic
systems\cite{ta01,ak98}.  Unlike in atomic systems, the range
and strength of the effective interparticle interaction among
the constituent particles are tunable over a wide range in
colloidal system. Hence, different structural orderings can be
realized with ease in these systems at ambient conditions .
Though the most dominant interparticle interaction among
like-charged particles in a charge stabilized colloidal
suspension is screened Coulomb repulsion, experimental
observations in bulk suspensions such as reentrant transition
\cite{ak88,ta96}, vapor-liquid transition\cite{ta96,ta92},
stable voids coexisting with ordered or disordered regions
\cite{ise88,ito94,ta97,ta98} have served as evidences for the
existence of a long-range attractive component in the effective
interparticle interaction  of like-charged colloids\cite{ta99}.
Purely a repulsive pair-potential given by
Derjaguin-Landau-Verwey-Overbeek(DLVO) theory \cite{der48} can
not explain the inhomogeneous nature or the phase separation
observed in like-charged colloidal suspensions. The DLVO theory
is advanced by  Sogami-Ise by considering the counterion
mediated attraction and regarding the macroionic part as  a
one-component system\cite{sog84}.  The effective pair-potential
$U(r)$ \cite{sog84} obtained from the Gibb's free energy is
found to have a long-range attractive term in addition to the
usual screened Coulomb repulsive term given by DLVO theory.
Tata et al's computer simulations using  this $U(r)$ could
explain the above mentioned experimental observations
satisfactorily
\cite{ta01,ta96,ta90,ta93,ta96a,ta98a}.

Recently there have been several alternative theories
\cite{van99,war00,chan01}  to understand the observed phase
separation in charged colloids.  These theories, which are also
known as "volume term theories" \cite{van99,war00} have strong
similarity in approach but differ  in technical details and are
the extensions of  Debye -Huckel theory for asymmetric
electrolytes.  These theories consider the total Helmholtz free
energy of the system that arises from Coulomb interaction
between all species in the system. The outcome of all these
theories is that pair-potential between colloidal particles can
remain repulsive as a function of interparticle distance $r$,
yet a gas-liquid or gas-solid phase separation can occur due to
volume terms which are responsible for the spinodal instability.
Recently Schmitz\cite{sch00} has brought out the connection
between volume term theories and the long-range attractive term
that  is present in Sogami-Ise potential $U(r)$.  It is shown
that Sogami-Ise theory includes  volume terms and the attractive
term in $U(r)$ arises from the electrostatic interactions that
are present in volume term theories. Further, the criticism  on
Sogami-Ise theory by Overbeek \cite{ob87} and
Woodward's\cite{wood88} has been countered by
Smalley\cite{sma90} and Schmitz\cite{sch93}. Hence, we use
$U(r)$ as the model potential in  our MC simulations.

The objective of this paper is two fold. (1) Carry out
simulations as a function of $\sigma$ for volume fractions and
salt concentrations which are the same as that used in
experiments of Yamanaka {\em et al} \cite{yam98} and look for
the reentrant solid-liquid phase behaviour observed by them .
The reentered disordered state observed at high values of
$\sigma$ is known to be inhomogeneous but the nature of the
underlying phase transition has not been identified. (2) To
identify conditions under which the reentered inhomogeneous
disordered state will transform to a homogeneous state. Towards
this end we  perform a detailed MC study over a wide range of
salt concentrations as $C_s$ is known to alter the position
$R_m$ of the minimum as well as the depth $U_m$ of $U(r)$. Our
simulations as a function of $\sigma$  show the reentrant
order-disorder transition in agreement with experimental
observations and we go on to  predict a inhomogeneous to
homogeneous transition as a function of $C_s$, that is yet to be
observed in charged colloidal suspensions.

The details of MC simulations are given in the Sec. II. The
effect of charge density on the phase behaviour of suspensions
with different volume  fractions at low salt concentrations are
discussed in Sec. III, where the suspensions exhibit reentrant
solid-liquid transition. Section IV includes the results on the
influence  of salt concentration on the reentered disordered
state.  Comparison of our simulation results with the results of
reported experiments is presented in Section V.  A brief summary
with conclusions are given in section VI.

\section{DETAILS OF SIMULATION}

Monte Carlo (MC) simulations are carried out using Metropolis
algorithm with periodic boundary conditions for a canonical
ensemble (constant $N$, $V$, $T$ where $N$, $V$ and $T$ are
respectively the number of particles, volume and temperature).
Particles of diameter $d$(=120nm) are assumed to interact via a
pair  potential $U(r)$ having the functional  form

\begin{equation}
U(r) = 2 \frac{(Ze)^2}{\epsilon} \left(\frac{\sinh (\kappa
d/2)}{\kappa d} \right)^2 \left(\frac{A}{r} -\kappa \right)
\exp(-\kappa r)
\end{equation}

\noindent
where $A = 2 + \kappa d$ coth($\kappa d/2$) and the inverse
Debye screening length $\kappa$ is given as

\begin{equation}
\kappa^{2} = 4\pi e^2 (n_pZ + C_s)/(\epsilon k_BT).
\end{equation}

\noindent $Ze$ is effective charge on the particle (related to the
surface charge  density by $\sigma = Ze/\pi d^2$),  $C_s$   the
salt concentration, $T$ the temperature(298K),  $\epsilon$   the
dielectric constant of water and $k_B$ is the Boltzmann
constant.  The position of the potential minimum $R_m$ is given
as $R_m = \{A +[A(A+4)]^{1/2}\}/2\kappa$ and its depth by $U_m =
U(R_m)$.  Both $R_m$ and $U_m$ depend on $\sigma$ and $C_s$. For
the required volume fraction $\phi$ (= $n_p \pi d^3/6$), the
length $l$ of the MC cell is fixed from the relation $l^3 =
N/n_p$. By performing simulation with different number of
particles, N, we found in our earlier simulations
\cite{ta96,ta93,ta98a} that the results are the same within the
statistical error for $N\ge$ 432. Hence N is fixed at 432.

The results presented in this paper correspond to $\phi$=0.005
and 0.03.  At these volume fractions the deionised suspensions
are known to crystallize into a body centered cubic (bcc)
ordering \cite{sir89}, hence we have chosen particles placed on
a bcc lattice as the initial configuration in all our
simulations.  We monitor the total interaction energy $U_T$ and
the first peak height of the structure factor, $S_{max}$ for
identifying thermal equilibrium unambiguously
\cite{bin79,ta92a}.  Most of the simulations away from the
transition (e.g., freezing) took approximately 8 $\times$ 10$^5$
configurations to reach equilibrium, while those close to the
transition needed nearly 6 $\times$ 10$^6$ configurations.  A MC
Step (MCS) is defined as $N$ attempted moves during which, on an
average, each particle gets a chance to move. The step size to
move the particles during MC evolution process is chosen in such
a way that the trial acceptance ratio is always around 50$\%$.
After reaching equilibrium  pair-correlation function $g(r)$,
coordinate averaged pair-correlation function $g_c(r)$ and
$S_{max}$ are calculated using procedures reported earlier
\cite{ta92a,ta91}.  $g_c(r)$ is obtained by averaging the
coordinates of the  particles over a sufficiently large number
of configurations\cite{ta91}.  Since $g_c(r)$ is free from
thermal broadening  and it helps in identifying  the crystal
structure unambiguously. Further, for suspensions which exhibit
solid-like behaviour, the $g_c(r)$ shows sharper peaks as
compared to the corresponding $g(r)$.  The mean square
displacement $<r^2(t)>$  for a  chosen particle is defined as
$<r^2(m)>$ = $<\mid {\bf r}_i(m+n)-{\bf r}_i(n)\mid^2>$, where
${\bf r}_i(m)$ is the position of $ i $ th particle after $m$
MCS and $<... >$ denotes the configurational average over the
initial configurations $n$. Cluster size and its distribution
are calculated as described in our earlier work\cite{ta93}. The
total fraction of particles participating in the clustering
$F_c$, defined as the ratio of the total number of particles
that participate in the clustering to the total number of
particles $N$, has also been obtained for suspensions which
exhibited inhomogeneous nature (phase separation).  We identify
the nature of suspension whether it is homogeneous or
inhomogeneous by calculating the ratio $d_s/d_0$, where  $d_0$
(= $\sqrt3/2[1/n_p]^{1/3}$) is the average interparticle
separation calculated from the particle concentration $n_p$ and
assuming a bcc-like coordination in the liquid-like order and
$d_s$ is the average interparticle distance estimated from the
first   peak position in $g(r)$. For a homogeneous suspension
$d_s/d_0$=1 and for an inhomogeneous suspension $d_s/d_0$ $<$ 1.

\section{EFFECT OF CHARGE DENSITY}

We have carried out MC simulations  for two volume fractions
$\phi$ = 0.005 and $\phi$ = 0.03  as a function of charge
density on the particles keeping the salt concentration at
2$\mu$M. These parameters are same as that used by Yamanka {\em
et al.}\cite{yam98}  in their experiments. Fig. 1 shows the
pair-correlation functions and projections of the corresponding
MC cell onto a $xy$-plane for suspensions with $\phi$=0.005 and
$\sigma$=0.1, 0.23 and 0.5 $\mu$C/cm$^2$. For low values of
$\sigma$ (=0.1 $\mu$C/cm$^2$),  $g(r)$ and $g_c(r)$ showed a
decay as a function of $r$,  the corresponding particle
positions(Fig.  1(B)) showed disorder and the ratio $d_s/d_0$ is
found to be one (see table I). These observations suggest the
suspension at this value of $\sigma$ is homogeneous and
liquid-like ordered.  On the other hand for suspension with
$\sigma$= 0.27 $\mu$C/cm$^{2}$ we find $g(r)$ does not decay
with $r$ and the positions of sharp peaks in $g_c(r)$ clearly
indicate that the ordering is bcc. The projection of  particles
in the MC cell  shows perfect ordering and the calculated
$d_s/d_0$ is found to be one confirming the bcc ordered
crystalline structure is homogeneous.  At high values of
$\sigma$ ( = 0.5 $\mu$C/cm$^{2}$)  $g(r)$ and $g_c(r)$  are
found to decay a function of $r$ once again .  Further, we
observe the first peak in $g(r)$ shifted to smaller $r$ implying
$d_s/d_0$ $<$ 1.  These observations suggest that suspension is
inhomogeneous and disordered.  The corresponding projected
particles in the MC cell also confirm this and showed particle
free regions (voids) coexisting with disordered dense region.
Since the $g_c(r)$ shows sharper features than the corresponding
$g(r)$, the dense disordered  region  is identified to be
solid-like (amorphous).

In order to identify the values of $\sigma$ at which the
homogeneous liquid (HL)  freezes into a homogeneous crystalline
(HC) order and then to an inhomogeneous disordered  (also
referred as a phase separated (PS)) state, the structural
parameter $S_{max}$ is calculated as a function of $\sigma$ and
is shown in Fig. 2.  The sudden increase in $S_{max}$ at
$\sigma$=0.23 $\mu$C/cm$^2$ (Fig. 2(a)) corresponds to the
freezing transition and the sudden drop in $S_{max}$ at
$\sigma$=0.33 $\mu$C/cm$^2$ corresponds to the transition from
homogeneous crystalline state to a phase separated (gas-solid
coexistence) state.  Simulations for suspension with $\phi$=0.03
showed similar phase behaviour  except that the corresponding
transitions have been found to occur at lower values of $\sigma$
(see Table I).

These observations  are  understood  from  the dependence of
$U(r)$  on $\sigma$.  At low charge density the suspensions
remain homogeneous because the particles which  are at a
distance $d_0$ will always experience screened Coulomb repulsive
interaction as $d_0 < R_m$ (see curve $a$ of Fig. 3).  The
structural ordering in the homogeneous state depends upon the
strength  of $U(r)$ at  $r$ = $d_0$. As the charge density on
the particles is increased the strength of repulsive interaction
increases. This results in freezing into a homogeneous
crystalline state. Since $R_m$ decreases monotonically  with
increase in $\sigma$, particles which are at separation of $d_0$
will experience attraction when $R_m \leq d_0$ (see curve $b$).
Due to strong attraction the particles condense into a
solid-like dense phase  leaving some particles in the rare phase
(gas-like) which appear  as voids when the fraction of the
volume occupied by the rare phase is smaller than that of the
dense phase.  Thus the occurrence of freezing of a homogeneous
liquid into a homogeneous crystalline state at lower values of
$\sigma$ and the system reentering into a disordered
inhomogeneous phase (gas-solid coexistence) at higher values of
$\sigma$ are understandable.

\section{EFFECT OF SALT ON INHOMOGENEOUS DISORDERED PHASE}

There have been reports of theoretical
\cite{ta98a,ta91,rob88,sen91} and experimental
investigations\cite{ak98,sir89,mon89} on the effect of $C_s$ on
melting/freezing of a homogeneous colloidal suspension. However,
the conditions under which an inhomogeneous disordered state,
which occurs in highly charged colloids, becomes homogeneous are
not known. When the well depth $U_m$ $<$ $k_BT$, the attraction
between particles becomes weak, hence the suspension  can remain
homogeneous.  It can be seen from Fig. 3 (curve $d$) that the
well depth $U_m$ decreases to values smaller than $k_BT$ at high
salt concentrations. So,  we performed MC simulations over a
wide range of salt concentrations keeping the other suspension
parameters fixed and results are summarized in Table II.   Fig.
4 shows the effect of increasing salt concentration on the
disordered dense phase which coexisted with voids at $C_s$ =
2$\mu$M (Fig.  1(c)).  For low values of $C_s$, $g(r)$ shows a
decay, but the structural correlations  persist to the full
length of the MC cell (Fig. 1(c)) suggesting that the disordered
dense phase is connected. The fraction of volume occupied by the
voids is found to be smaller than that occupied by the dense
phase.  Hence the voids constitute the minority phase at low
values of $C_s$. The particle  positions in the MC cell(see Fig.
1(c)) confirm this.  When the salt concentration is increased we
observe structural correlations in $g(r)$ (Fig.  4(a)) to extend
up to  a few times the diameter of the particle. The first peak
height is also found to be very high and the first peak position
matches exactly with $R_m$. These observations imply the
formation of dense phase clusters with $R_m$ as the  average
distance between particles  within the cluster. The other sharp
peaks appearing at distances beyond the first peak corresponds
to short-range ordering of  particles within the clusters
(intracluster ordering).

Since the clustering occurs at large values of $C_s$, the
ordering of particles within a cluster could be liquid-like or
solid-like (amorphous). In order to distinguish between the two
types of ordering we have calculated $g_c(r)$,  g(r)  and the
mean square displacement $<r^2(t)>$ of particle within  the
cluster.  We compare the behaviour of these quantities with that
corresponding to a vapor-liquid coexistence state.  This
vapor-liquid coexistence state obtained by   performing
simulations for suspension parameters at which a vapor-liquid
coexistence has been reported \cite{ta01,ta96,ta92}.  It can be
seen from Fig. 5 that the peaks in $g_c(r)$ (curve $a$)
corresponding to $C_s$=550$\mu$M are much sharper with increased
peak height as compared to the corresponding $g(r)$ (Fig. 4(a)).
Whereas the $g_c(r)$ corresponding to the vapor-liquid
coexistence (dotted curve in Fig. 5) is much lower in peak
height and also poorer in  features as compared to the
corresponding g(r) (inset of Fig. 5). The mean square
displacement of particles inside the cluster corresponding to
$C_s$=550$\mu$M shows a saturation behaviour as a function of MC
time and undergo much smaller displacement as compared to that
of particle within a liquid-like ordered cluster. These
observations unambiguously suggest that the clusters formed   at
higher values of $C_s$ for highly charged colloids are
solid-like.  These clusters can be observed either in the
scattering or microscopy experiments provided the density of
solvent is matched with the density of colloidal particles. Tata
{\em et al} have observed the coexistence of voids with dense
phase disordered regions \cite{ta97,ta98}  and existence of
clusters under density matched conditions \cite{ta98} in highly
charged polycholorostyrene sulfonate particles (PCSS) dispersed
in an aqueous medium.  Density matching for PCSS particles has
been achieved by redispersing them in 60\%  glycerol aqueous
solution.

Simulations at a  salt concentration  of 1000 $\mu$M showed only
a single peak in g(r) (see Fig. 4(b)) at $r =R_m$.  The
corresponding  projection of particles in the MC cell (see inset
of Fig. 4 (b)) do not show clustering and are distributed
uniformly. These observations suggest that  the suspension with
$C_s$= 1000 $\mu$M is homogeneous and noninteracting (gas-like).
Thus we observe for $C_s$ $\leq$ 550 $\mu$M an inhomogeneous
state in the form of dense phase clusters having solid-like
ordering and a homogeneous gas-like disordered state at
$C_s$=1000$\mu$M. Hence the inhomogeneous (PS) to homogenous
disordered transition is expected to occur in between these salt
concentrations.  We identify this transition by monitoring the
total fraction of particles participating in the clustering
$F_c$ (Fig. 6(a)) and the first peak height $g_{max}$ of $g(r)$
(Fig. 6 (b)).  Note the sudden change in both of these
parameters around $C_s$ = 570$\mu$M as $C_s$ is increased. This
change is associated with the inhomogeneous (PS) to homogenous
(HG) transition or in other words transition from a gas-solid
coexistence to a gas phase. The large values of $g_{max}$ in the
inhomogeneous state arises due to scaling of $g(r)$ with respect
to particle concentration $n_p$.  The  dense phase concentration
$n_d$ is estimated from the first peak position ( $\cong$ $R_m$)
of $g(r)$ and is found to be several times higher than $n_p$
(see Table II). In the case of suspension with $\phi$=0.03 the
transition is found to occur at $C_s$=270$\mu$M which is lower
than $C_s$ = 570$\mu$M for $\phi$ = 0.005.  Suspensions of
higher volume fraction undergo PS to HG transition  at lower
values  of $C_s$. This is due to increased screening  of
particles in suspensions of higher volume fraction as compared
to that in suspensions with lower $\phi$.

\section{COMPARISON WITH EXPERIMENTS}

It is important to  compare our simulations results with that
observed experimentally. Our simulations as a function of
$\sigma$ revealed the existence of the three phases {\em viz.,}
a homogenous liquid (HL) at low $\sigma$, a homogenous
crystalline (HC) state at intermediate values of $\sigma$ and an
inhomogeneous disordered state (PS) at high values of $\sigma$.
Further, we showed that the homogenous liquid freezes into a
homogeneous crystalline state at about $\sigma$ $\ge$ 0.22
$\mu$C/cm$^2$ and disorders once again   at about $\sigma$ $ge$
0.33 $\mu$C/cm$^2$. Yamanaka {\em et al.}\cite{yam98} have
reported freezing of a homogenous liquid-like ordered aqueous
suspension of charged particles with $\phi$=0.005 into a
homogenous crystalline state for charge densities beyond 0.24
$\mu$C/cm$^2$.  The suspension is found to disorder once again
when $\sigma$ is increased beyond 0.40 $\mu$C/cm$^2$.  This
reentered disordered state is identified to be inhomogeneous
from the scattering measurements. The measured average
interparticle separation $d_{exp}$ is found to be less than
$d_0$\cite{yam98,yam96}, which implies the nonspacefilling
nature of the suspension at high charge densities.   Thus
present simulations are in good agreement with the  experimental
observations. Further, our simulations at higher volume fraction
have revealed that the values of $\sigma$ at which  HL to HC
transition and  HC to PS transition occur are lower as compared
to that in suspensions of lower $\phi$. This result is also in
agreement with experimental observations\cite{yam98}.

Present simulations on the reentrant inhomogeneous disordered
state as a function of salt concentration revealed that the
inhomogeneous state undergoes a transition to homogeneous gas
phase beyond  a critical salt concentration. This gas-solid
transition which can be observed  in highly charged colloids by
varying salt concentration, has not yet investigated
experimentally.  Further, we notice from these simulations the
absence of vapor-liquid coexistence for highly charged colloids.
In this context, it is worth mentioning the result from MC
simulations on a system of hard particles with an attractive
Yukawa interaction by Hagen and Frankel \cite{ha94}. They showed
the disappearance of vapor-liquid coexistence when the range of
attractive part of the Yukawa potential is less than
approximately one sixth of hard-core diameter. In present
simulations the range of attractive and repulsive parts of
$U(r)$ have been varied by changing $\sigma$ and $C_s$.  The
range of attraction is found to be smaller for suspensions of
high charge density particles as compared to the suspensions of
low charge density particles.  Hence, one can observe a
vapor-liquid coexistence in suspension of low charge density
particles \cite{ta96,ta92} but not in suspensions of highly
charged particles.

\section{CONCLUSIONS}

MC simulations using a pair-potential with long-range attractive
term  show a homogenous liquid, homogenous crystal and a
disordered inhomogeneous phase as the charge density is varied,
suggesting a reentrant order-disorder transition,  which is in
agreement with the observed experimental results. The reentered
inhomogeneous phase is identified as the gas-solid coexistence
state and occurs due to strong attraction experienced by the
highly charged particles.  Suspensions exhibit homogeneous
liquid-like and homogenous crystalline orders at lower values of
$\sigma$ due to the dominant  screened Coulomb repulsion at
average interparticle separation. The inhomogeneous disordered
state undergoes a transition to a homogenous gas state upon
increasing the salt concentration. This transition is analogous
to the gas-solid transition observed  in atomic and
colloid-polymer mixtures as a function of temperature and free
polymer, respectively\cite{ile95}. We  believe that our
prediction of gas-solid transition driven by the salt
concentration would motivate experimentalists to investigate
highly charged colloids in search of this transition as well as
for detailed phase diagrams with a gas-liquid or gas-solid
coexisting phases.

\noindent
Authors wish to thank M.C. Valsakumar for valuable discussions
and helpful suggestions.

\noindent $^\ast${Author to whom correspondence should be
addressed.}

\newpage

\begin{figure}
\caption{(A) $g(r)$ vs  $r$ for different charge densities at a
salt concentration of 2$\mu$M and $\phi$ = 0.005.  Curves  $a$,
$b$ and $c$ correspond  to $\sigma$ = 0.1, 0.23, 0.50
$\mu$C/cm$^2$  respectively. The inset shows $g_c$(r) vs $r$.
The curves $a'$, $b'$, $c'$ correspond to the same  parameters
as those of curves $a$, $b$ and $C$. Curves $b$, $c$,  $b'$ and
$c'$ are shifted vertically for the sake of clarity. (B)
Projection of time-averaged particle coordinates in the MC cell
for parameters same as $a$, $b$ and $c$.}

\caption{Structural parameter $S_{max}$ as  a  function  of
$\sigma$ for a suspension with $C_s$ =2.0 $\mu$M  and $\phi$
=0.005. The line drawn through the points is a guide to the
eye.}

\caption{Pair potential $U(r)$/$k_BT$ for different suspension
parameters  with $\phi$=0.005. Curve $a$ and  $b$ correspond to
$\sigma$= 0.15 and 0.5 $\mu$C/cm$^2$, respectively and $C_s$=2.0
$\mu$M.  Curve $C$ and $d$ correspond to  $C_s$=550.0 and 1000
$\mu$M, respectively and $\sigma$ = 0.15 $\mu$C/cm$^2$. The
vertical line corresponds to the average interparticle
separation $d_0$.}

\caption{ $g(r)$ vs $r$  at two different salt concentrations (a)  $C_s$=
550.0 $\mu$M and  (b) at $C_s$= 1000.0 $\mu$M. Insets show  the projection
of particles in the  MC cells corresponding to the two values of $C_s$.  Other parameters
of the suspension are  $\phi$ =0.005, $\sigma$ =0.5
$\mu$C/cm$^2$.}

\caption{(a) $g_c(r)$ vs $r$ for  suspensions exhibiting gas-solid
(full curve: $\phi$= 0.005, $\sigma$= 0.5 $\mu$C/cm$^2$) and
vapor-liquid (dotted line: $\phi$= 9.02 x 10$^{-3}$, $\sigma$=
0.21$\mu$C/cm$^2$, $C_s$=40.8$\mu$M, $d$=0.109$\mu$m) coexistence.
$g(r)$ vs $r$ for the vapor-liquid coexistence is shown as inset.
(b) Mean square displacement $<r^2(t)>$ vs $t$ (in units of 1000
MCS)  for particles within a solid-like  (full curve) and
liquid-like (dashed curve) clusters.}

\caption{Quantities  (a) $F_c$  and (b) $g_{max}$   as  a
function of $C_s$ used for the identification of the transition from
a inhomogeneous disordered state to homogenous gas phase. Other
parameters of the suspension are same as that of Fig. 4.}
\end{figure}


\begin{table}
\caption{Inverse  Debye  screening length ($\kappa$), position
of  the potential minimum ($R_m$), depth  of the  potential well
($U_m$), average interparticle separation ($d_0$) and average
interparticle separation $d_s$ calculated from first peak
position of $g(r)$  for different values of $\sigma$
corresponding to  two suspensions with widely differing
volume fractions.  The abbreviations  HL, HC and PS represent
the homogeneous liquid, homogeneous crystalline  and  phase
separated  states, respectively.}
\begin{tabular}{ccccccccc}

$\phi$ & $\sigma$ ($\mu$ C/cm$^2$) & $\kappa d$ &
$R_m/d$ & $U_m/k_BT$ & $d_0/d$ & $d_s/d_0$ & State  \\ \hline

0.005& 0.065 & 0.667 & 7.350 & -0.088 & 5.143 & 1.0 & HL\\
0.005& 0.100 & 0.718 & 6.841 & -0.221 & 5.143 & 1.0 & HL\\
0.005& 0.150 & 0.786 & 6.270 & -0.539 & 5.143 & 1.0 & HL\\
0.005& 0.200 & 0.848 & 5.837 & -1.023 & 5.143 & 1.0 & HL\\
0.005& 0.230 & 0.883 & 5.600 & -1.399 & 5.143 & 1.0 & HC\\
0.005& 0.273 & 0.930 & 5.340 & -2.050 & 5.143 & 1.0 & HC\\
0.005& 0.300 & 0.960 & 5.190 & -2.547 & 5.143 & 1.0 & HC\\
0.005& 0.410 & 1.070 & 4.690 & -5.173 & 5.143 & 0.9 & PS\\
0.005& 0.500 & 1.152 & 4.383 & -8.119 & 5.143 & 0.8 & PS\\
0.030& 0.060 & 1.021 & 4.890 & -0.107 & 2.830 & 1.0 & HL\\
0.030& 0.150 & 1.462 & 3.542 & -0.850 & 2.830 & 1.0 & HC\\
0.030& 0.200 & 1.660 & 3.100 & -1.608 & 2.830 & 1.0 & HC\\
0.030& 0.273 & 1.900 & 2.841 & -3.142 & 2.830 & 1.0 & HC\\
0.030& 0.300 & 1.990 & 2.744 & -3.853 & 2.830 & 1.0 & PS\\
0.030& 0.410 & 2.300 & 2.460 & -7.348 & 2.830 & 0.9 & PS\\
0.030& 0.500 & 2.530 & 2.300 & -10.90 & 2.830 & 0.8 & PS\\ \hline
\end{tabular}
\end{table}

\newpage
\begin{table}
\caption{Inverse  Debye  screening length ($\kappa$), position
of the potential minimum ($R_m$), depth  of the   potential well
($U_m$), and the ratio of dense phase concentration $n_d$ to
homogeneous particle concentration $n_p$ for different values of
$C_s$ for suspensions with volume fractions $\phi$=0.005 and
0.03 exhibiting inhomogeneous and homogeneous states.  The
abbreviations PS and HG  represent the phase separated state
and a homogeneous gas  state,  respectively.}
\begin{tabular}{cccccccc}

$\phi$ & $\sigma$ ($\mu$ C/cm$^2$) &  $C_s$ ($\mu$M) & $\kappa d$ & $R_m/d$
 & $U_m/k_BT$ & $n_d/n_p$  & State  \\ \hline

0.005 & 0.5 &  10    & 1.607  & 3.270 & -9.909 & 3.890  & PS\\
0.005 & 0.5 &  20    & 2.037  & 2.700 & -10.76 & 6.980  & PS\\
0.005 & 0.5 &  50    & 2.975  & 2.070 & -10.50 & 15.33 & PS\\
0.005 & 0.5 & 100   & 4.086  & 1.740 & -8.541 & 27.11 & PS\\
0.005 & 0.5 & 200   & 5.688  & 1.520 & -5.757 & 39.04 & PS\\
0.005 & 0.5 & 300   & 6.930  & 1.422 & -4.267 & 49.71 & PS\\
0.005 & 0.5 & 350   & 7.474  & 1.390 & -3.774 & 53.19 & PS\\
0.005 & 0.5 & 400   & 7.980  & 1.360 & -3.382 & 55.96 & PS\\
0.005 & 0.5 & 500   & 8.908  & 1.328 & -2.803 & 60.72 & PS\\
0.005 & 0.5 & 550   & 9.340  & 1.313 & -2.583 & 63.54 & PS\\
0.005 & 0.5 & 600   & 9.743  & 1.300 & -2.395 & 1.000 & HG\\
0.005 & 0.5 & 700   & 10.521& 1.278 & -2.095 & 1.000 & HG\\
0.005 & 0.5 & 800   & 11.241& 1.261 & -1.860 & 1.000 & HG\\
0.005 & 0.5 & 900   & 11.918& 1.246 & -1.670 & 1.000 & HG\\
0.005 & 0.5 & 1000 & 12.558& 1.234 & -1.524 & 1.000 & HG\\
0.030 & 0.3 & 10     & 2.285  & 2.474 & -3.933 & 1.500 & PS\\
0.030 & 0.3 & 50     & 3.389  & 1.915 & -3.548 & 3.230 & PS\\
0.030 & 0.3 & 100   & 4.395  & 1.680 & -2.858 & 4.782 & PS\\
0.030 & 0.3 & 150   & 5.211  & 1.565 & -2.335 & 5.916 & PS\\
0.030 & 0.3 & 200   & 5.915  & 1.495 & -1.958 & 6.786 & PS\\
0.030 & 0.3 & 250   & 6.544  & 1.446 & -1.680 & 7.500 & PS\\
0.030 & 0.3 & 300   & 7.117  & 1.410 & -1.471 & 1.000 & HG\\
0.030 & 0.3 & 400   & 8.145  & 1.358 & -1.176 & 1.000 & HG\\
0.030 & 0.3 & 500   & 9.055  & 1.323 & -0.980 & 1.000 & HG\\
0.030 & 0.3 & 600   & 9.883  & 1.296 & -0.841 & 1.000 & HG\\ \hline
\end{tabular}
\end{table}
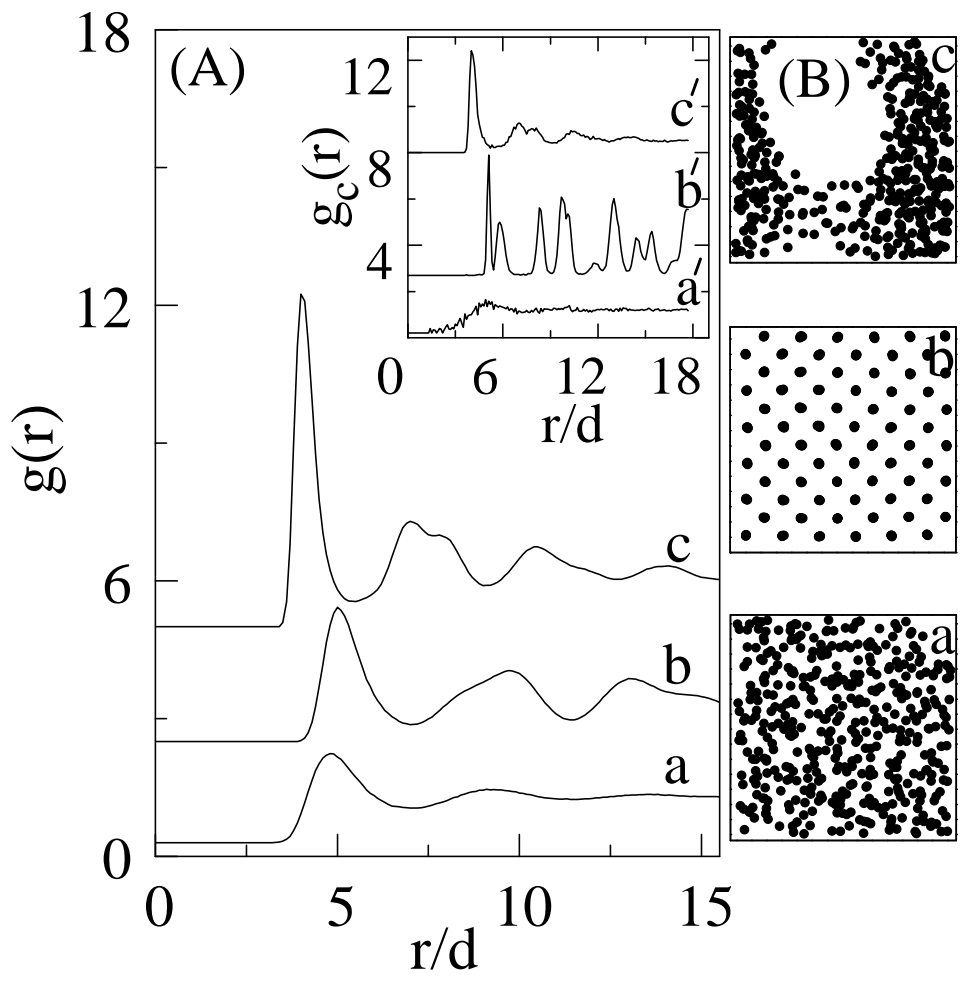
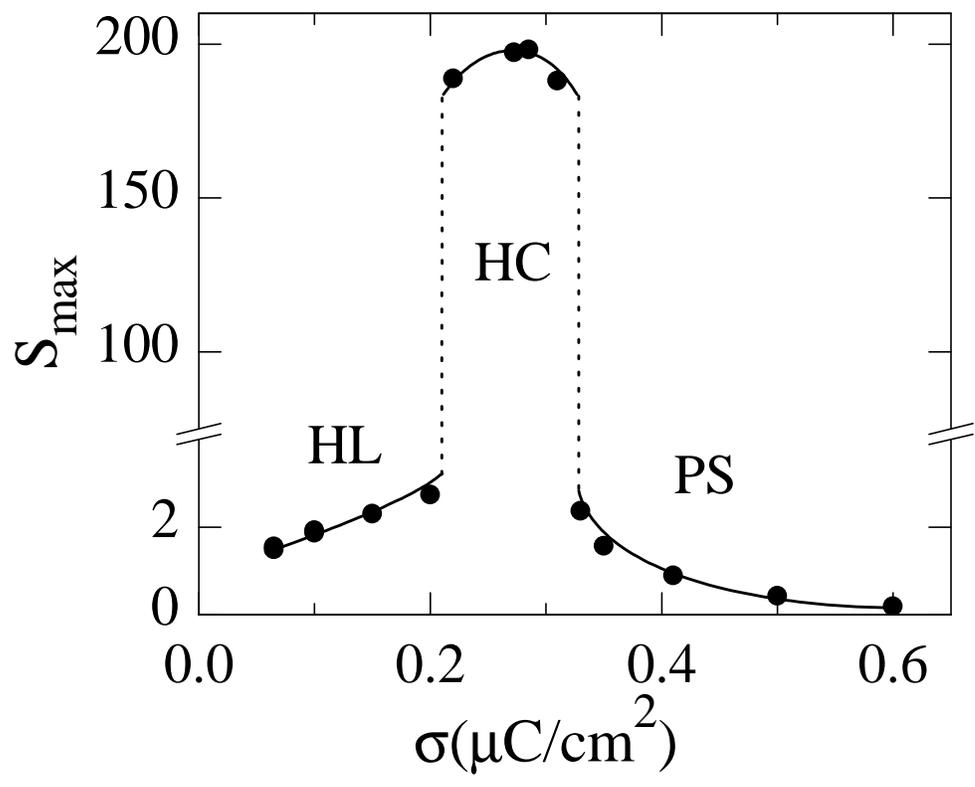
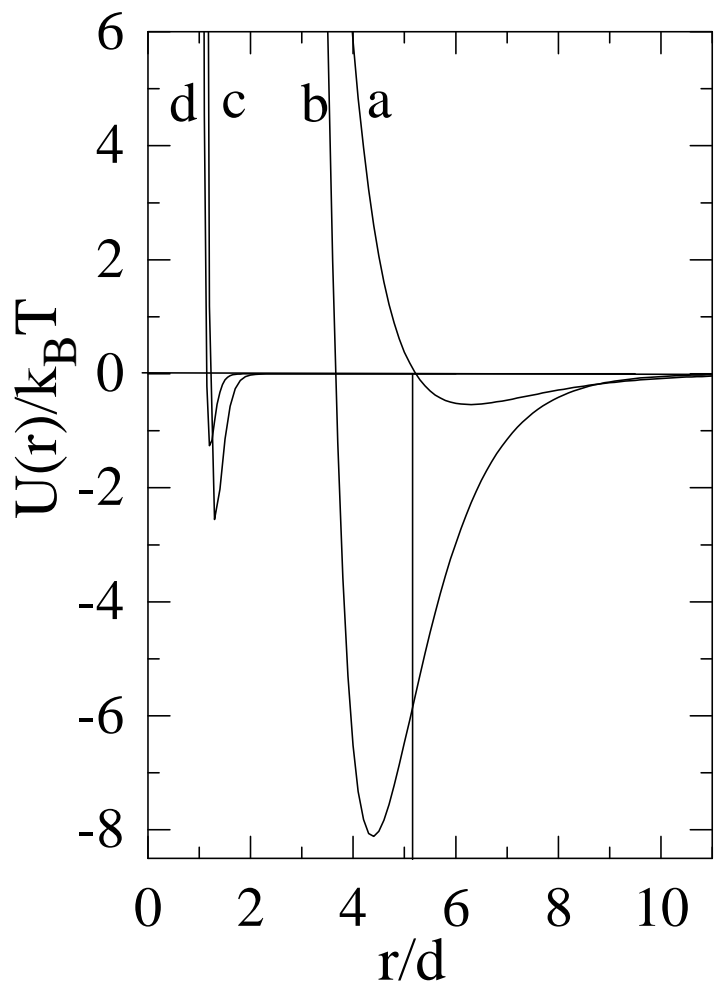
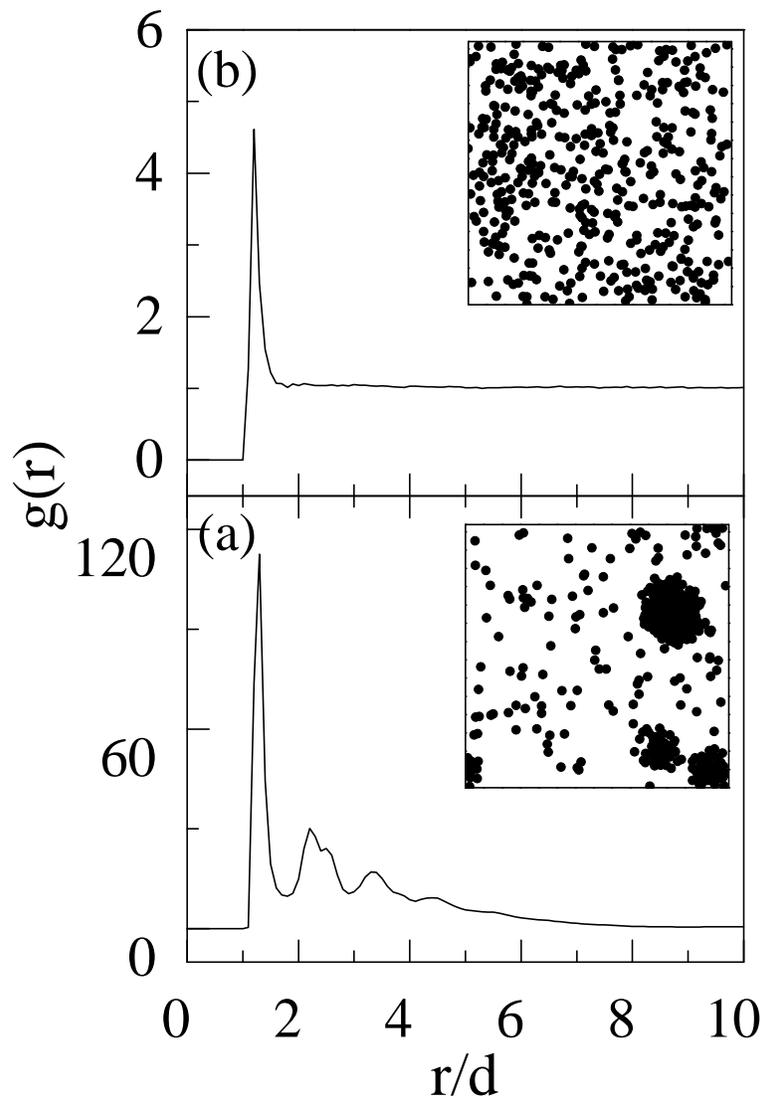
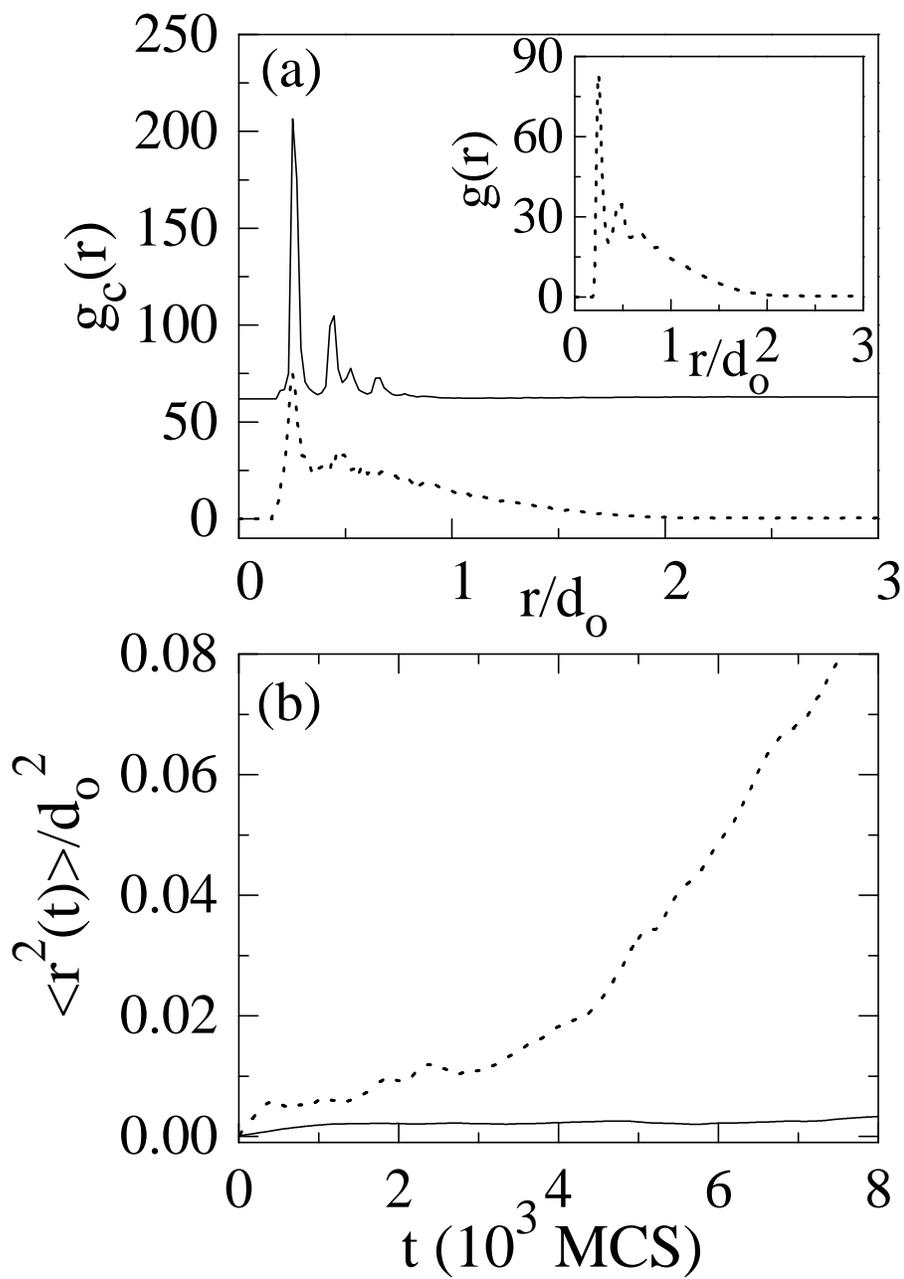
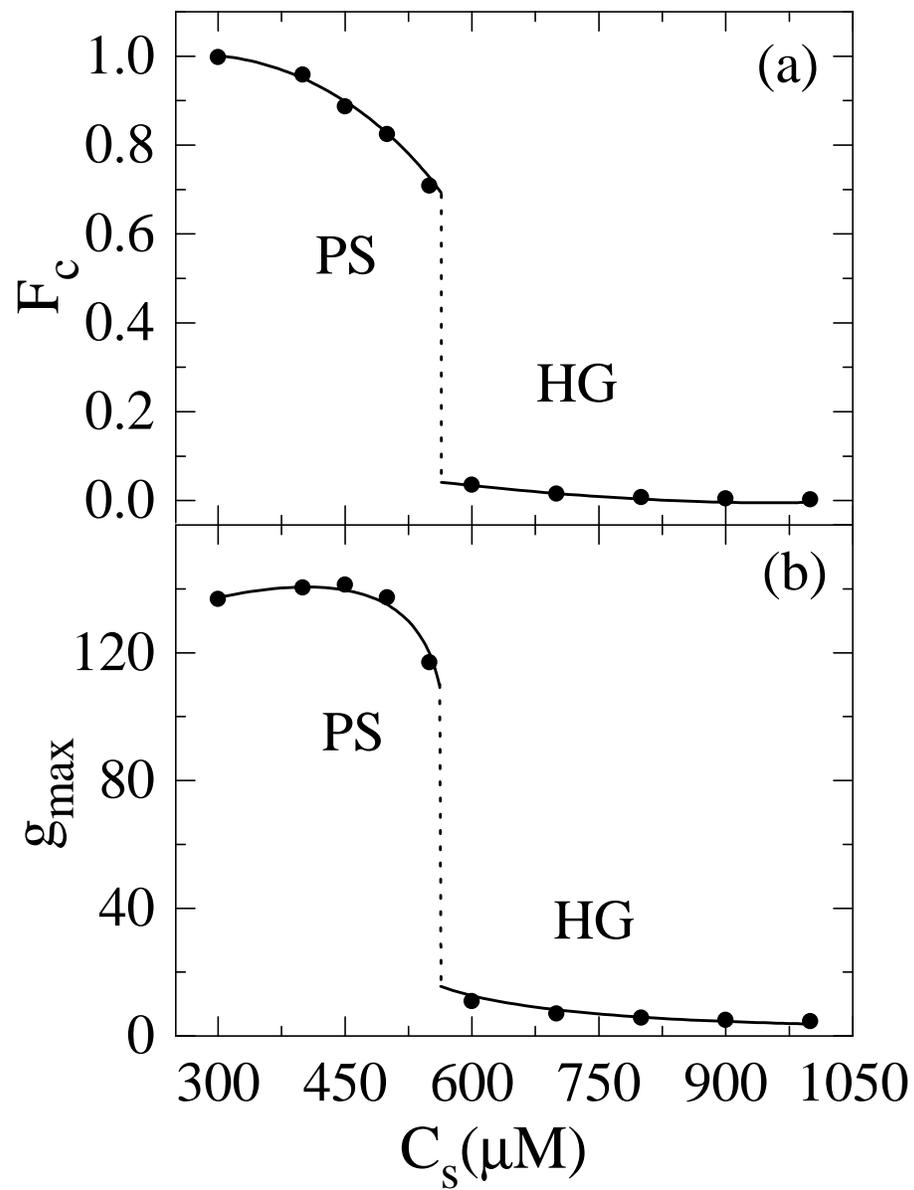
\end{document}